\documentclass[floatfix,aps,prl,10pt,twocolumn,a4paper,amsmath,amssymb]{revtex4}

\usepackage{graphicx}
\usepackage{bm}
\def\real{{\bf R}}
\usepackage{times}
\begin{document}

\newcommand{\be}{\begin{eqnarray}}
\newcommand{\ee}{\end{eqnarray}}

\title{
A quantitative analysis of concepts and semantic structure in written language:
Long range correlations
in dynamics of texts}
\author{E. \'Alvarez-Lacalle}
\affiliation{Dept.~of Complex Systems,
Weizmann Institute of Science, Rehovot, Israel}
\author{B. Dorow}
\affiliation {Institute for Natural Language Processing, University
of Stuttgart, Germany}
\author{J.-P. Eckmann}
\affiliation {Dept.~de Physique Th\'eorique et Section de
     Math\'ematiques, Universit\'e de Gen\`eve, Switzerland}
\author{E. Moses}
\affiliation{Dept.~of Complex Systems, Weizmann Institute of
Science, Rehovot, Israel}
\date{\today}

\begin{abstract}
Understanding texts requires memory: the reader has to keep in mind
enough words to create meaning. This calls for a relation
between the memory of the reader and the
structure of the text.
To investigate this interaction, we first identify
a connectivity matrix defined by
co-occurrence of words in the text.
A vector space of words characterizing the text is spanned
by the principal directions of this matrix.
It is useful to think of these weighted
combinations of words as representing ``concepts''. As the reader
follows the text, the set of words in her window of attention
follows a dynamical motion among these concepts. We observe long
range power law correlations in this trajectory. By explicitly
constructing  surrogate hierarchical texts, we demonstrate that
the power law originates from structural organization of texts
into subunits such as chapters and paragraphs.

\end{abstract}
\maketitle

\section{Introduction}
Language is a central link through which we interact with other people.
As a channel of communication it is
limited by our physical ability to speak only one word at a time.
The question arises therefore how the complex products of
our brain are transformed into the linear string of words that
comprise speech or text. Since our mental processes are far from
being one dimensional, the use of memory is essential, as is the
existence of some type of correlations in time.

Such questions have a long and intense history. Bolzano
\cite{Bolzano} already noted on the need for specific organization in
scientific texts, while Ingarden devotes the book \cite{Ingarden} to
understanding the process by which a text is understood and
assimilated. Modern methods \cite{528470,manning-foundations}
combine the work of linguists with those of computer scientists,
physicists, physiologists and researchers from many other fields
to cover a wide range of texts: from the phoneme
\cite{Clark/Yallop:1995}, going
on to words
\cite{widdows-geometry,yarowsky95unsupervised,pereira-distributional,Dor03,Beate2004}
and grammar \cite{charniak97statistical,929278}, all the way to
global text analysis \cite{beeferman-segmentation}
and the evolution of language
\cite{Nowak,dorogovtsev01languageAs}.

Recent interest has focused on applying methods of statistical physics to
identify possible trends and correlations in text
\cite{Peng,SZZ,Shnerb,MZ}.
In \cite{MZ}, for example, the authors study the distribution of
words across different works by the
same authors, combining notions of information, entropy and statistics to
define a random walk on the text. Long ranged correlations have been found in
the writings of Shakespeare and Dickens, and a number of hypotheses as to their
origin have been proposed. These include the overall existence of ideas and
meaning \cite{SZZ,Shnerb} or of some semantic hierarchy \cite{MZ}.

Here we aim at a middle ground both in methods of analysis and in
ranges of text, based on geometric intuition developed in
\cite{EM2002,EMS2004}.
We use a variant of Latent Semantic Analysis
\cite{deerwester-indexing,schutze-automatic,ferrer01theSmall}  to
uncover semantic connections in the text. This method leads to the
identification of cohesive groups
of words, and we find it useful to think of these groups as describing
``concepts''.  The most important of these groups are singled out,
and form an (orthogonal) basis of a vector space defined by the
text.
We then introduce a dynamics on the space of concepts
and find long range power-law
correlations in time. We end by showing that the origin of these
correlations can be found in the hierarchical structure
\cite{Barabasi2003} of the text. In this way, we are able to connect
with the classic work of Bolzano and Ingarden on the
intelligibility of texts.

\section{The concept space}

We define on the set $\{w\}$ of words in the text an arbitrary
order, $w_1,\ w_2,\dots$, for example using their rank, which is
the number of times $m_i$ that word $w_i$ appears in the text
under consideration. We associate with $w_1$ the vector
$(1,0,0,\dots)$, with $w_2$ the vector $(0,1,0,\dots)$, and with
$w_i$ the vector which has a ``1'' at position $i$ and all others
``0''. These vectors span a vector space in which each axis
corresponds to one single word.

The analysis proceeds with the construction of a symmetric
connectivity matrix $M$ based on co-occurrence of words. This
matrix has rows and columns indexed by words, and the entry
$M_{ij}$ counts how often word $w_i$ and $w_j$ co-occur within a
distance $d$ in a given text. We suggestively term $d$ as the
``window of context'', and typically take it to be of size
$100$.

The connectivity matrix $M$ is then normalized to take into account the
abundance of words. If the $m_i$ occurrences of $w_i$ are randomly
distributed and are not closer to each other than $2d$ (a reasonable
assumption if $d\ll L$) then the probability that any of the
occurrences $m_j$ of word $w_j$ will randomly fall within a distance
$d$ them
is given by
$$
R_{ij}\,=\,\frac{2d m_i m_j }{L}
$$
so that $R$ is the connectivity matrix of the corresponding ``random
book'', with $d$ the context window defined before
and $L$ the number of words in the book.
The normalized connectivity
matrix is then
\begin{equation}\label{e:1}
N_{ij}\,=\, \frac{M_{ij} -R_{ij}}{\sqrt{R_{ij}}}~.
\end{equation}
This normalization quantifies the extent to
which the analyzed text deviates from a random book (with the same
words) measured in units of its standard deviations.

To improve the statistical significance, as well as cut the matrix
down to a manageable size, we only consider words that occur enough
times in the book.
We define a threshold value $m_{\rm thr}$, which the
number of
occurrences $m_i$ must exceed for word $w_i$ to be included.
$m_{\rm thr}$ is set by the random normalization $R_{ij}$, and must
therefore be proportional to $\sqrt{L/d}$. We found empirically that
$m_{\rm thr} \geq 0.3 \sqrt{{L}/{d}}$ gave consistently good
statistical significance.

Discarding words with lower $m_i$ reduces the effect of single
co-occurrences between rare words,
where Eq.~(\ref{e:1}) would lead to unrealistically high $N_{ij}$ ($\gtrsim2$).
In the texts we considered, the values of the cut-off range from
$m_{\rm thr}=4$ to $23$,
(see Table~\ref{tab:values}). Words that cross this threshold are
``significant" and are indexed from $i=1$ to $P$.

\begin{table}
\begin{tabular}{|c|r|r|r|r |r|c|}
\hline
Book & Length & $m_{\rm thr}$ & P & T  & $S_{\rm conv}$ & Exponent \\
\hline
MT &  22375 & 4~  & 342 & 16.2  & 25~ &  0.45 (0.05) \\
HM &  32564 & 5~  & 425 & 15.5  & 30~ &  0.92 (0.03) \\
NK &  62190 & 8~  & 750 & 20.4  & 60~ &  0.81 (0.03) \\
TS &  73291 & 8~  & 644 & 16.8  & 45~ &  0.50 (0.03) \\
DC &  77728 & 8~  & 781 & 19.8 & 75~  &  0.43 (0.07) \\
IL & 152400 & 12~  & 774 & 22.0 & 75~ &  0.40 (0.05) \\
MD & 213682 & 14~  & 1162 & 20.0  & 75~ &  0.45 (0.04) \\
QJ & 402870 & 20~  & 1246 & 18.7  & 100~ &  0.36 (0.03) \\
WP & 529547 & 23~  & 1498 & 22.9  & 300~ &  0.43 (0.03) \\
EI &  30715 & 5~  & 448 & 25.3  & 50~ &  0.75 (0.10)\\
RP & 118661 & 11~  & 609 & 14.5  & 75~ &  0.60 (0.06)\\
KT & 197802 & 14~  & 661 & 25.4  & 50~ &  0.30 (0.05) \\

\hline
\end{tabular}
\caption{Table of book parameters and results. $m_{\rm thr}$ is the threshold
for the number of occurrences and $P$ is the number of words kept
after thresholding. $T$ is
the percentage of the words in the book that pass the threshold, 
$T=\frac{\sum^P _{i=1} m_{\rm i}}{L}$. $S_{\rm conv}$ is the dimension
at which a power law is being fit. The exponent of the fit
is given in the last column, together with its error in parenthesis.}
\label{tab:values}
\end{table}

Once we have reduced the size of the matrix $N$, we change the basis by
performing a singular value decomposition (SVD), and keeping only
those $S$ vectors with highest singular values. We will use the
terminology of rectangular matrices, even in the case of square
(symmetric) ones, as we are going to use later matrices with
unequal numbers of rows and columns. We therefore use the terms
singular vector rather than eigenvector and singular value rather
than eigenvalue. We will discuss below the effect of varying the
truncation parameter $S$.

The idea behind this choice of principal directions is that the
most important vectors in this decomposition (those with highest
singular value) describe {\em concepts}. A connectivity matrix
similar to the one we use has been introduced before
\cite{pantel02document,Dor03}, based on adjacency of words rather than our
looser requirement that words appear together within a wider
window. This resulted in the ability to cluster words according to
context and identify ambiguity in words \cite{Beate2004}. What we
derive here may be viewed as a large-scale version of the
assignation of meaning by co-occurrence, in comparison to the
local result obtained previously \cite{Beate2004}.

We now have a basis of the SVD such that every word can be
described as a unique superposition of the basis vectors.
Thus,
$$
e_i = \sum_{j=1}^S a_{ij} v_j~,
$$
where $e_i$ is the vector of all zeros except at position $i$
(representing the word $w_i$) while the $v_j$ are the vectors of
the SVD.

\section{Texts}

We used twelve books---in their English version---for our
analysis. Nine of them are novels: ``War and Peace"  by Tolstoi
(WP),``Don Quixote" by Cervantes (QJ), ``The Iliade" by Homer
(IL), ``Moby Dick" (MD) by Melville, ``David Crocket" by Abbott
(DC), ``The Adventures of Tom Sawyer" by Twain (TS), ``Naked
Lunch" by Burroughs (NK), ``Hamlet" by Shakespeare (HM) and
``Metamorphosis" by Kafka (MT). They span a variety of periods,
styles and also have very different lengths (see Table~\ref{tab:values}).

Besides the nine novels we analyzed the scientific didactic book
``Special theory of relativity " by Einstein (EI), the
philosophical treatises ``The Critique of Pure Reason" by Kant (KT)
and ``The Republic" by Plato (RP).

Each of the books is processed by eliminating punctuation and
extracting the words. Each word is ``stemmed'' by querying the
on-line database WordNet 2.0, and the leading word for this query
is retained without keeping the information on whether it was
originally a noun, a verb or an adjective .


All the stop words, {\it i.e.}, words that carry no significant
meaning, are assigned a value of zero. The list of these words
consists of  determiners, pronouns, and the like. This
standard list is supplemented with a list of
broadly used words which are abundant
independently of the text.
In practice we reject those words whose occurrences are above $m_{\rm thr}$
in over $90\%$ of the books.

Books are thus transformed into a list of stemmed words with which the
connectivity matrix is defined, and to which the SVD process is applied.

\begin{table}
\begin{tabular}{|l|l|l|l|l|l|}
\hline
    MD(1) & MD(5) & EI(1) & EI(2) &TS(1) &TS(2)\\
\hline
whale & bed & 	 surface & planet & spunk & ticket \\
ahab & room & Euclidean & sun & 	 wart & bible \\ starbuck &
queequeg & being & ellipse & huck & verse \\
sperm & dat & 	 universe & mercury & nigger & blue \\ cry & 	 aye
& 	 rod & 	 orbital & reckon & yellow \\ aye & 	 moby &
	 spherical & orbit & stump & pupil \\ sir & 	 dick & plane
& star & bet & 	 ten \\ boat & landlord & geometry & arc &
midnight & spunk \\ stubb & ahab & continuum & angle & johnny & red \\
leviathan & whale & sphere & second & em & thousand \\
\hline
\end{tabular}
\caption{Examples of the highest singular components for three books.
Given are component one and five of Moby Dick (MD), one and two of
Einstein (EI) and of Tom Sawyer (TS). The coefficients of the words
in the singular component may be positive or negative and their
absolute values range from $0.13$ to $0.3$.}
\label{tab:words}\end{table}

Examples of the concept vectors from the different books are
illuminating (see Table~\ref{tab:words}). The
first ten words in the principal component with highest singular value
in Moby Dick immediately carry us into the frame of the story and
introduce many of its protagonists. The next three principal
components are somewhat similar, with the addition of familiar words such as
white, shark, captain, ship. By the fifth largest principal component a
change of scene occurs as the story takes a detour indoors,
and this is evidenced by the second column of Table~\ref{tab:words}.

Similarly, the first ten words of the principal component with highest
singular value of Einstein's ``Special relativity" launch us
immediately into the subject matter of special relativity, while its
second component brings in the applications to astrophysics.
It is perhaps amusing to recall the tales of Tom Sawyer by viewing the
principal component with highest singular value. These deal with Tom's
various escapades, for example the bible competition which Tom wins
by procuring tickets by various trades and bargains.

We can conclude that the ``concepts'' we defined by using singular vectors
do indeed capture much of the content of the text.

\section{Dynamic analysis}

Having found a representative basis for each of the texts, our
main interest is in the dynamics of reading through the text.
What is new here in comparison with
earlier statistical analysis \cite{MZ} or linguistic
research \cite{Brunet74}
is that the
basic ingredient is not the byte (as in the statistical studies)
nor the word, but rather a contextual collection of words (our
concept vector). In this way, our study links the word
connectivity matrix to semantic meaning.

\begin{figure}[h]
\label{TOM}
\includegraphics[width=80mm]{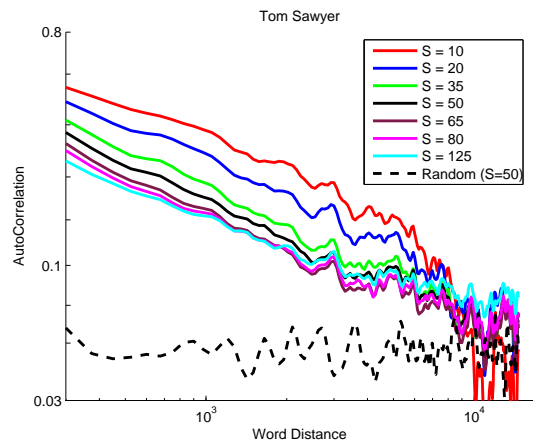}
\caption{Log-log plot of the autocorrelation function for Tom
Sawyer using different numbers of singular components for building
the dynamics. For comparison, the autocorrelation of a randomized
version of the book is also shown. \label{Fig.1}}
\end{figure}

Basically, we again slide a ``window of attention'' of fixed size $A$
along the text
and observe how the corresponding vector moves in the vector space
spanned by the SVD. If this vector space is irrelevant to the
text, then the trajectory defined in this space would probably be
completely stochastic, and would perform a random walk. However,
if the evolution of the text is reflected in this vector space,
then the trajectory should trace out the concepts alluded to
earlier in a systematic way, and some evidence of this will be
observed.

To keep the algorithm reasonably simple, we divide the length of the
text into $L/A$ non-overlapping windows, where a value of $A=150$
words is a good choice. We can gain some intuition in this vector
space by replacing the notion of distance along the text (measured in
segments of $A$ words) with the concept of time (measured by the time
it takes a hypothetical reader to read $A$ words). We define time as
$t=\ell\times\delta t$, with
$\ell$ the index of the window and $\delta t$ the average time it takes
to read $A$ words.
For each window we obtain a vector $V(t)$
which is decomposed as:
$$
V(t)=\sum_{j=1}^S a_j(t)v_j~,
$$
with the SVD basis $\{v_j\}$ chosen as before.

The moving vector $V(t)\in\real^S$ is a dynamical system and
we proceed to study its autocorrelation function in time
$C(\tau) = \langle V(t)V(t+\tau)\rangle_t -\langle
V(t)\rangle_t^2$, where $\langle \cdot \rangle _t$ is
the time average.
Fig.~\ref{Fig.1} shows
the correlation function of the concept vector in time for ``Tom
Sawyer" given in a log-log scale. The different lines correspond
to different values of $S$. The function is non-zero over a large
range, on the order of more than $10^3$ words. This range is much
longer than what we found when measuring correlations among
sentences, without using the concept vectors (data not shown).

As the dimension $S$ increases the correlation function
in the log-log representation converges to a straight line,
indicating a very clear power law behavior. The convergence to a
power-law behavior and the dimension necessary to produce it
depend on the book.

The range at
which the correlation is significant and above the noise depends
both on the exponent and on the natural noise in the system. The
noise in turn depends both on the quality of the expansion in
terms of the SVD and on the cohesiveness of the text.

\begin{figure}[h]
\label{Short}
\includegraphics[width=80mm]{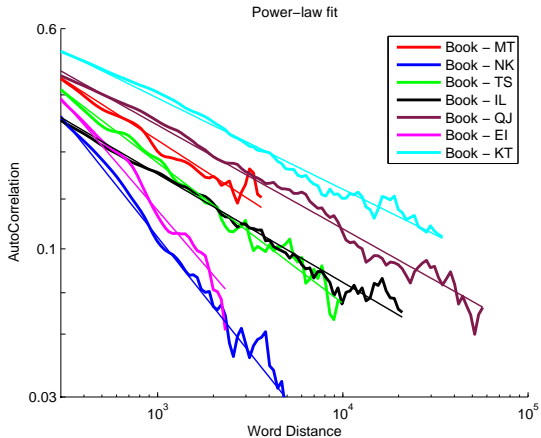}
\caption{Autocorrelation functions and fits for seven of the books
listed. The autocorrelation functions are truncated at the level where
the noise sets in.
\label{Fig.2}}
\end{figure}

The results presented in
Table~\ref{tab:values} (see Fig.~\ref{Fig.2}) are given for the lowest value of
$S$ at which the convergence to a power-law behavior is clearly discerned.

The long range correlations uncovered in this fashion are in line
with previous measurements obtained using the random walk approach
of \cite{Peng, Shnerb, MZ}. However, the range over
which we find correlations is much larger, and the quality of the
power law fit is accordingly significantly better.

\section{Controls}

The methods we have described above require a certain number of
parameters, such as the threshold rank value $m_{\rm
thr}$ of the matrix, or the size of the windows that are being
moved along the text. We describe here some tests which were
performed to check the robustness of the method when these
parameters are changed, summarizing the most relevant findings:

\begin{table}
\begin{tabular}{|c|c|r|c|r|c| c|}
    \hline Book & $m_{\rm row}$ & $m_{\rm col}$ & $P_{\rm rows}$ &
$P_{\rm col}$ &  T & Exponent \\
\hline
MT & 1 & 4~  & 1709 & 342  & 24.9  &  0.65 (0.05) \\
HM & 1 & 5~  & 3440 & 425  & 29.4  &   1.40 (0.20) \\
NK & 2 & 8~  & 3665 & 750  & 35.8 &  1.00 (0.10) \\
TS & 2 & 8~  & 2748 & 644  & 26.4 &  0.70 (0.10) \\
DC & 2 & 8~  & 3071 & 781  & 30.1  &  0.60 (0.08)\\
IL & 3 & 12~  & 2392 & 774  & 27.8  &  0.48 (0.05)\\
MD & 4 & 14~ & 4007 & 1162 & 29.1  &  0.50 (0.05) \\
QJ & 5 & 20~ & 3574 & 1246 & 24.2  & 0.45 (0.03)  \\
WP & 6 & 23~ & 4130 & 1498 & 28.6  &  0.50 (0.05)  \\
EI & 1 & 5~  & 1721 & 448  & 32.0  &  1.00 (0.20)\\
RP & 3 & 11~  & 2020 & 609  & 20.5  &    0.70 (0.05)\\
KT & 4 & 14~ & 1549 & 661  & 28.6  &  0.37 (0.03) \\

\hline
\end{tabular}
\caption{Book parameters and results for rectangular connectivity
matrices. The values of $S_{\rm ms}$ are the same as in
Table~\ref{tab:values}.
The threshold value for the rows
$m_{\rm row}$  is smaller than that of the columns $m_{\rm col}$ so
accordingly the number of rows $P_{\rm rows}$ is bigger than the number
of columns $P_{\rm col}$.
$T$ is as defined in Table~\ref{tab:values}.  The
dynamics change when more words are added and thus the
exponents are larger, that is, the correlations are weaker.}
\label{tab:valuesII}
\end{table}

1) The threshold  must be chosen carefully. By
lowering the threshold $m_{\rm thr}$ of accepted words, and therefore increasing 
the
number of accepted words, one observes a systematic decrease of
correlations. One can
take a matrix of $P_{\rm rows}$ rows and $P_{\rm col}<P_{\rm rows}$
columns.
The results are shown in Table~\ref{tab:valuesII} and should be
compared to those of Table~\ref{tab:values}.

2) A change in the size of the window of attention (the variable $A$)
does not affect the results significantly as long as it is
kept above $100$ words and not much bigger than
the window of context ($d=200$ words).
A lower value of $A$ means a lower number
of words per window and a correspondingly higher noise.

3) We checked, to some extent, the language dependence
of the method,
by comparing ``Don Quixote'' in Spanish and English.
While languages can have quite different local syntactic rules,
the long term correlations practically do not depend on the
language. This is perhaps related to the importance
of nouns in creating the correlation function, and
these are translated more or less one for one.

\section{Hierarchy and the origin of scaling}

The existence of power laws is often traced to hierarchical
structures \cite{Barabasi2003}. We put forward the hypothesis that in our
case these structures are parts of the texts (such as ``volumes'',
``parts'', ``chapters'' within parts, ``sections'' in chapters,
``paragraphs'' in sections, and so on). This is a hierarchy of $K$
levels, each containing several parts. For example, a book may be in
$3$ volumes that each have about $10$ chapters, each of which is
divided in $8$ or so sections, etc. For simplicity, we assume that
each level contains the same number of parts $H$.
Typical values are $K=4$ and $H\approx 7$. The important point is
that the text has the structure of a tree.

We now show that the power law we found earlier for the text is
not changed if words are permuted in the text, provided one
respects as much as possible the structure of the book as a whole.
As discussed above, if the structure is not kept, the randomized
text has no correlations.

We prepare an (initially empty) hierarchical book as a template
into which we will insert the words from the original book. The
empty book is divided in $H$ roughly equal parts, each subdivided
again in $H$ roughly equal parts. This subdivision is repeated $K$
times. We end up with the book divided in $K$ levels and a total
of $H^K$ subdivisions at the smallest scale. $K$ and $H$ are
chosen so that the lowest level corresponding to ``paragraph''
will have around $100$ words.

\begin{figure}[h]
\label{Long}
\includegraphics[width=80mm]{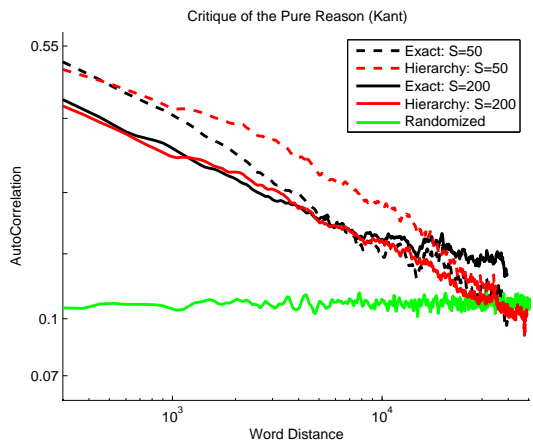}
\caption{Comparison of autocorrelation functions for the original
book of Kant (black), the randomly re-organized version (green)
and the hierarchically reorganized version (red) using $E=5$.
\label{Fig.3}}
\end{figure}

We place each word into the hierarchical book individually.
Assume the word $w_i$ appears $m_i$ times in the original text.
Fix a parameter $E> 1$ (for concreteness take $E=5$). We define
weights recursively for each subdivision. At the top level, each
part has the initial weight $J$. Choose one part randomly and
confer on it weight $J\cdot E$. The next level inherits the
weights introduced before. Now repeat the choice of a random part
from the second level, and multiply its weight by $E$. Depending
on which slot has been chosen, at the second level there may be
one slot with weight $JE^2$, and the others have weight $JE$ or
$J$, or there are weights  $JE$ and $J$ only. Going on in this
fashion we fill all the levels, reaching finally a range of
weights $JE^k$, with $k\in\{0,\dots,K\}$ at the lowest level. We
then choose $J$ so that the sum of weights is one, and distribute
the $m_i$ copies of word $w_i$ randomly according the weights in
the finest subdivision. This procedure is applied to all words and
produces a hierarchical randomized text, that preserves the word
distribution and resembles the structural hierarchy of the book.

As seen in Fig.~\ref{Fig.3}, performing this hierarchical
randomization process on
the book preserves the power law behavior of correlations found in the
original text. Since the simple randomization destroys the
power law see Fig.~\ref{Fig.3} we can conclude that the power
laws of the original text do indeed originate in their hierarchical
structure.

One can improve the fit of the power law by introducing
further parameters: For example, skipping randomly some levels in
the construction some of the weights $JE^k$ gives (this
corresponds to pruning some branches of the tree) or by changing
the value of $E$ for each word, specifically , by increasing the
value of $E$ for  some words with lower rank (this comes from the
observation than lower ranks words tend to be more concentrated
around certain paragraphs or chapters than higher rank words).

\section{Conclusions}

Many questions remain to be addressed, for example applying the
dynamic approach to spoken text, in which repetitions are known to be
of importance, and comparing the results to those of written text. It
may also be of interest to characterize different types of text or of
authors according to the correlation exponent. It remains to be seen
whether the hierarchical organization we have identified in texts is
related to a hierarchical organization in our thought processes.

Our approach enables the quantification and rigorous examination
of concepts that have been introduced long ago and discussed
heuristically by the great classics in the field. Bolzano, in his
Wissenschaftslehre, written in 1837, studies the theory of
scientific writing, and points out in great detail how such
writing should proceed. In particular, in Vol III, he points out
that, starting from ``symbols'' (he probably thinks of
Mathematics) one works one's way to a fully structured text,
containing paragraphs, sections, chapters, and so on. He clearly
instructs the reader of how to maintain the intelligibility of the
text. Ingarden, in his ``vom Erkennen des literarischen
Kunstwerks'' talks, from his philosopher point of view about the
activity of the brain which compresses parts of texts so that they
may be more easily recalled. The entities he has in mind are
``layers of understanding'' ( 16, p.111 : \dots not every layer of
an already read part of a text is kept in the same way in
memory, \dots The reader keeps bigger and smaller
text-connections---{\em Satzzusammenh\"ange}---in his living
memory \dots )

Our study allows to measure the degree to which the insights of
authors like these can be understood. It adds therefore a new piece to
the puzzle of understanding the nature of language.

\bibliography{refs}

\end{document}